\documentclass[runningheads]{llncs}
\usepackage[T1]{fontenc}
\usepackage{graphicx}
\usepackage{amsmath}
\usepackage{amssymb}
\usepackage{bbm} 
\usepackage{microtype} 
\usepackage{hyperref}
\usepackage{bbding}
\usepackage{cite}
\usepackage{color}

\begin{document}
\title{Advancing Sentence-Level Dysarthric Audio-Visual Speech Recognition via Test-Time Adaptation}
\titlerunning{Sentence-Level DAVSR via Test-Time Adaptation}
\author{
Shiyao Wang\orcidID{0009-0006-5786-7109} \and 
Hui Wang\orcidID{0009-0003-8057-4644} \and 
Jiaming Zhou\orcidID{0009-0002-4819-4572} \and 
Shiwan Zhao\orcidID{0000-0001-5068-025X} \and 
Yong Qin\orcidID{0009-0000-2748-3020}\inst{*}\Envelope
}
\renewcommand{\thefootnote}{}
\footnotetext{* Corresponding author.}
\authorrunning{S. Wang et al.}
\institute{
Academy for Advanced Interdisciplinary Studies, Nankai University, Tianjin, China\\
\email{\{wangshiyao, wanghui\_hlt, zhoujiaming\}@mail.nankai.edu.cn},\\
\email{zhaosw@gmail.com}, \email{qinyong@nankai.edu.cn}
}
\maketitle   
\begin{abstract}
Dysarthric audio-visual speech recognition (DAVSR) enhances performance by integrating complementary visual cues. However, existing research is largely confined to word-level tasks with limited vocabularies, which severely restricts real-world communication for dysarthric speakers. Furthermore, performance remains heavily dependent on supervised fine-tuning with speaker-specific data; significant variations in age, etiology, and severity cause traditional methods to struggle with unseen speakers. To address these domain shifts and the scarcity of continuous multimodal resources, we first introduce a "source speech search-segmentation" pipeline to construct a structured sentence-level corpus from raw CDSD videos. We then propose the use of unsupervised Test-Time Adaptation (TTA) for sentence-level DAVSR using Sequential-level Generalized Entropy Minimization (SGEM). Our evaluations across the state-of-the-art AV-SUPERB and ViSpeR frameworks demonstrate that TTA consistently alleviates Character Error Rate (CER) degradation for unseen pathological speakers. Paired $t$-tests verify that these improvements are statistically significant ($p < 0.05$), highlighting TTA as a promising paradigm for advancing speaker-adaptive assistive speech interfaces.

\end{abstract}
\renewcommand{\thefootnote}{\arabic{footnote}} 
\section{Introduction}
Dysarthric speech recognition \cite{pbdsr_ori,pblrdwws_ori,dda,sapc_ori} is a vital catalyst for social integration, directly enhancing the quality of life for individuals with speech disorders by enabling voice-command interaction with smart home devices. By extending standard audio-visual speech recognition (AVSR) \cite{autoavsr_ori}, DAVSR \cite{davsr20191_ori,davsr20192_ori,davsr2020_ori,davsr2023_ori,davsr20241_ori,davsr20242_ori} leverages multi-modal cues---integrating audio signals with visual data such as lip movements---to provide essential robustness when articulatory impairments compromise audio clarity. Recent advancements include the use of Bayesian networks to predict bimodal uncertainty \cite{davsr20191_ori} and gated neural networks for dynamic modal fusion \cite{davsr20192_ori}. Regarding speaker-specific optimization, researchers have employed speaker-dependent Learning Hidden Unit Contribution (LHUC) \cite{lhuc_ori} for fine-tuned adaptation \cite{davsr2020_ori}, while neural codec language modeling has been applied to reconstruct dysarthric speech into intelligible, typical forms \cite{davsr20242_ori}.

Despite the potential of DAVSR, research on sentence-level tasks remains underdeveloped. Most existing studies are confined to word-level recognition \cite{uaspeech_ori} characterized by limited vocabularies and speaker-specific training. In practice, progress is hindered by two primary challenges: \textbf{data scarcity} and \textbf{task complexity}. Regarding the former, publicly accessible dysarthric corpora are generally small-scale and word-oriented; sentence-level resources are rare and often lack the precise segmentation required for robust modeling. While the recent CDSD dataset \cite{cdsd_arxiv_ori} helps mitigate this shortage, its release as raw, unprocessed video necessitates extensive pre-processing before it can be effectively utilized for model training. Furthermore, the inherent heterogeneity of dysarthric speech---driven by diverse ages, etiologies, and severity levels---complicates continuous speech modeling and severely limits generalization to unseen speakers. Conventional pipelines typically rely on target-speaker fine-tuning to bridge this pronunciation gap; however, this strategy is both data-intensive and brittle when faced with shifts in speech severity or varying recording conditions \cite{var_ori}.

\begin{figure}[t]
\setlength{\abovecaptionskip}{0.1cm} 
\centerline{\includegraphics[width=10cm]{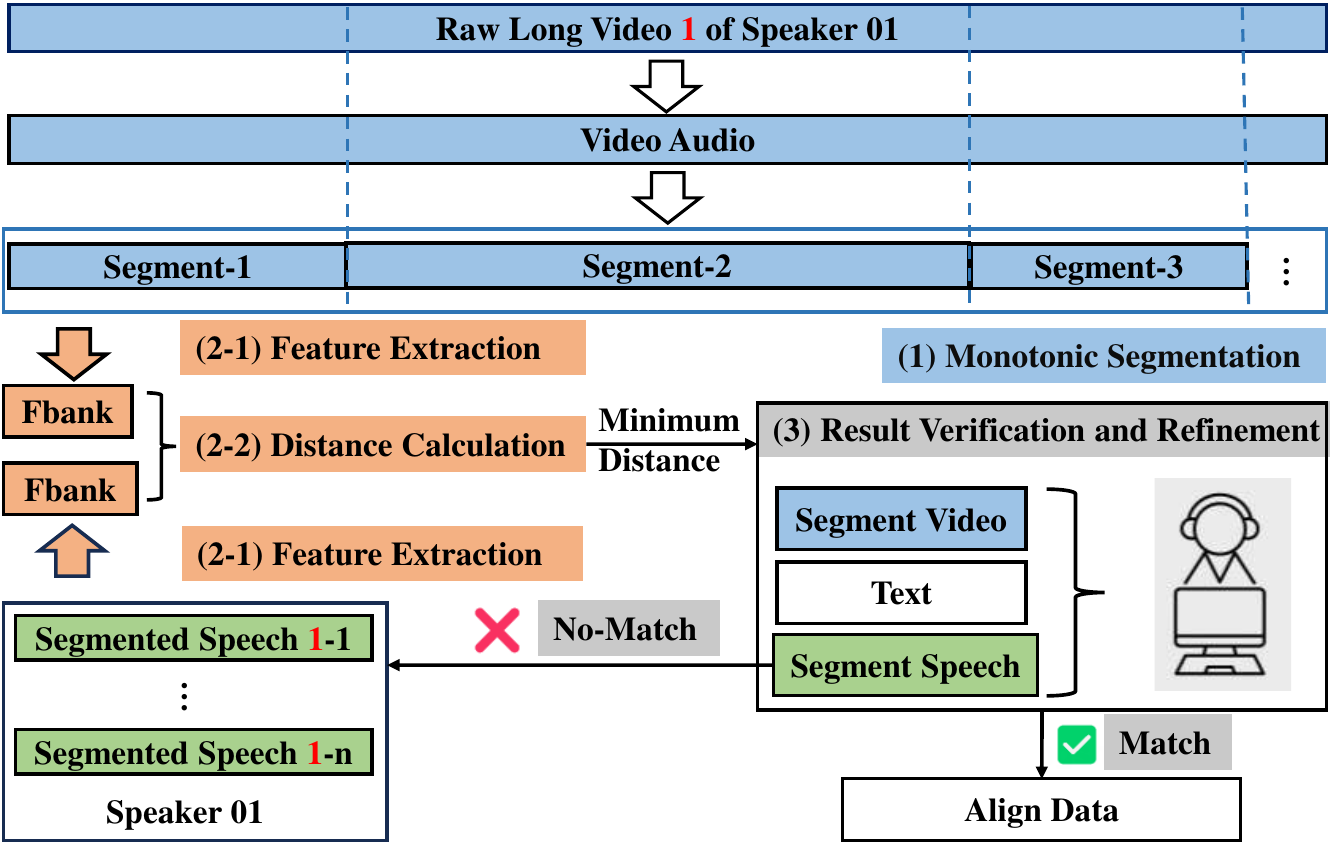}}
\caption{The workflow of the source speech search-segmentation procedure. The framework processes raw audio-visual streams through three core stages: (1) monotonic segmentation to ensure chronological consistency; (2) Fbank feature extraction and distance calculation to locate candidates via minimum Euclidean distance; and (3) multi-modal manual result verification and refinement to filter out no-match cases and finalize the aligned sentence-level corpus.}
\label{fig:ssss}
\end{figure}

To address data scarcity, we introduce a \emph{source speech search-segmentation} procedure (Figure \ref{fig:ssss}) that automatically aligns and extracts sentence-level units from raw CDSD videos. This pipeline bridges the data-preparation gap, laying the groundwork for the systematic study of continuous DAVSR. To address task complexity, we explore \emph{Test-Time Adaptation} (TTA) as an unsupervised mechanism to cope with speaker variability and domain mismatch. TTA adapts models on unlabeled test samples at inference, reducing the discrepancy between training and deployment without requiring labeled target data. While TTA has shown promise in image classification \cite{tent_ori,sita_ori} and unimodal ASR \cite{SUTA_ori,SGEM_ori}, its potential in the multi-modal DAVSR domain---where labeled data are scarce and robustness to unseen speakers is critical---remains largely unexplored.

The main \textbf{contributions} of this work are summarized as follows:
\begin{itemize}
\item We shift the focus of DAVSR research to sentence-level tasks, moving beyond restricted word-level vocabularies toward real-world communication scenarios.
\item We develop a ``source speech search-segmentation" method to transform raw CDSD video data into structured, sentence-level segments, providing a crucial resource for the research community.
\item We provide the first systematic validation of TTA for DAVSR across multiple state-of-the-art frameworks (AV-SUPERB\cite{avsuperb_ori} and ViSpeR\cite{ViSpeR_ori}), exploring its feasibility and demonstrating potential performance improvements for unseen dysarthric speakers under various configurations.
\end{itemize}

\section{Methods}
\subsection{Source Speech Search-Segmentation}
\label{sec:Source speech search-segmentation}
To construct a viable sentence-level DAVSR corpus, we must align manually annotated speech segments with the unsegmented raw videos in the CDSD dataset. We propose a "source speech search-segmentation" method (Fig.\ref{fig:ssss}) designed to ensure precise one-to-one correspondence while preventing alignment redundancy. The procedure consists of three core stages:

\subsubsection{Monotonic Segmentation.} We implement a monotonic matching mechanism to enforce temporal consistency. Specifically, once a speech segment is aligned with a video clip, all subsequent segments are restricted to the remaining, unmatched portion of the video. This sequential constraint inherently prevents repeated matches and ensures that the audio-visual pairs maintain their original chronological order, significantly enhancing the reliability of the alignment pipeline.

\subsubsection{Feature Extraction and Distance Calculation.} We first extract Filter bank (Fbank) features from the pre-segmented speech utterances. To find the corresponding match, we slide a window across the audio stream of the raw video to extract candidate segments of the same duration as the target utterance and compute their Fbank features. The alignment is then formulated as a distance minimization problem: we calculate the Euclidean distance between the feature set of the target utterance and those of all candidate video segments, selecting the segment that yields the minimum distance as the optimal match.

\subsubsection{Result Verification and Refinement.} To ensure high data fidelity, we conduct rigorous manual verification of the initial alignment results. For utterances where the initial alignment fails or is deemed inaccurate, the search procedure initiates an iterative refinement by re-traversing the entire raw video up to three times to locate the correct segment. If a valid match remains elusive after these exhaustive traversals, the sample is excluded from the final dataset to maintain corpus integrity. To guarantee the reliability of this alignment pipeline, the "Result Verification and Refinement" stage explicitly involved four independent human annotators. An audio-visual segment was validated and finalized as an official "Match" only upon achieving unanimous consensus across the acoustic, textual, and visual modalities. This rigorous cross-validation and manual filtering protocol effectively eliminates subjective bias, ensuring the absolute structural integrity and fidelity of the resulting sentence-level alignment.

\begin{table}[t] 
    \setlength{\abovecaptionskip}{0.0cm} 
    \caption{Statistics of the CDSD dataset.}
    \vspace{-6pt}
    \label{tab:cdsd data} 
    \centering
    \begin{tabular}{l | c c | c c | c | c}
        \hline
        \textbf{ID} & \multicolumn{2}{c|}{\textbf{Segmented Speech}} & \multicolumn{2}{c|}{\textbf{Raw Video}} & \textbf{Aligned} & \textbf{Availability} \\ 
        & \textbf{Count} & \textbf{Range (s)} & \textbf{Count} & \textbf{Range (s)} & \textbf{Data} & \textbf{Rate}\\
        \hline
        \textbf{01} & 1107 & [0.5, 10.8]& 22 & [136.2, 279.6] & 1107  & 100.0\% \\
        \textbf{02} & 988 & [0.4, 12.6] & 12 & [237.6, 348.6] & 968 & 98.0\%\\
        \textbf{03} & 926 & [0.4, 15.7] & 12 & [154.8, 585.6] & 912 & 98.5\% \\
        \textbf{04} & 1035 & [0.5, 10.4] & 14 & [252.6, 307.2] & 1035 & 100.0\% \\
        \textbf{06} & 969 & [0.6, 9.2] & 9 & [297.6, 744.0] & 967 & 99.8\%\\
        \textbf{10} & 1129 & [0.4, 7.7] & 18 & [144.0, 382.2] & 1101 & 97.5\% \\
        \textbf{22} & 1866 & [0.4, 5.6] & 16 & [61.2, 299.4] & 1866 & 100.0\% \\
        \textbf{44} & 1323 & [0.4, 4.9] & 2 & [393.6, 487.2] & 294 & 22.2\%\\  
    \hline
    \end{tabular}
\end{table}

Table \ref{tab:cdsd data} summarizes the CDSD dataset statistics before and after this procedure. We report the quantity and duration of speech segments, raw video metrics, and the final count of validated audio-visual pairs. We evaluate the method using the Availability Rate, defined as the ratio of successfully matched segments to the total pre-segmented count. Notably, all subjects except Speaker 44 achieved an Availability Rate exceeding 97\%. The lower yield for Speaker 44 is attributed to highly ambiguous pronunciation and minimal labial movement, which impeded both automated matching and manual verification. Conversely, the clear visemic cues provided by other speakers facilitated a highly efficient alignment. Furthermore, even when a speaker fails to articulate a specific phoneme due to impaired motor control of their vocal organs, human annotators can successfully verify and match the segment by cross-referencing visual information—a capability that embodies the foundational rationale for integrating visual cues into AVSR tasks. These results demonstrate the efficacy of our method in resolving the CDSD alignment challenge, providing a robust foundation for sentence-level DAVSR. The segmentation scripts will be open-sourced to the research community.

\begin{figure}[t]
\setlength{\abovecaptionskip}{0.0cm} 
\centerline{\includegraphics[width=12.5cm]{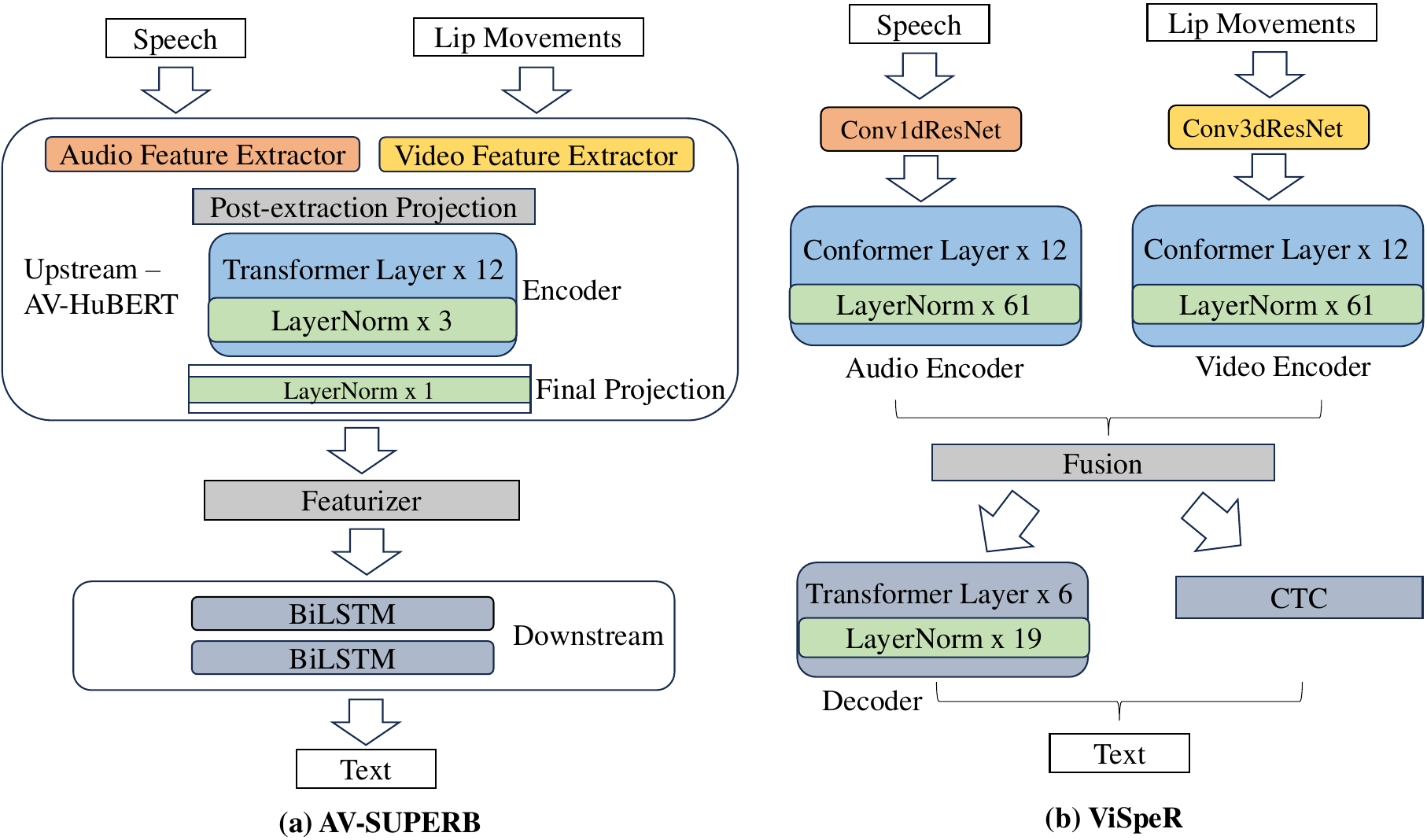}}
\caption{TTA was evaluated on two audio-visual speech recognition frameworks: AV-SUPERB and ViSpeR. Colors indicate different TTA setups, identifying the audio feature extractor, video feature extractor, cross-modal fusion layer, encoder, layer normalization (LayerNorm) layers, and decoder, respectively. Color viewing is recommended.}
\label{fig:tta-f3}
\end{figure}

\subsection{Architecture and Methodology}
\label{sec:avsr}
To establish a comprehensive evaluation setup, two mainstream AVSR frameworks with distinct multi-modal processing strategies—AV-SUPERB and ViSpeR—are utilized as our experimental testbeds, with their detailed architectural layouts systematically illustrated in Fig.~\ref{fig:tta-f3}. Built upon these distinct baseline infrastructures, we introduce TTA into the recognition pipeline to enhance DAVSR performance for unseen speakers. By leveraging unlabeled test data during inference, the TTA paradigm operates by dynamically modulating specific trainable components within these architectures. This design effectively mitigates the severe domain shifts and speaker variability inherent in dysarthric speech, allowing the frameworks to unsupervisedly adapt to unseen pathological speech characteristics while fully preserving their native architectural advantages.

\subsubsection{AV-SUPERB.} This framework follows a "pre-trained representation + downstream adaptation" paradigm. It utilizes a pre-trained AV-HuBERT \cite{avhubert_ori} backbone as the upstream encoder to extract robust multi-modal features. Specifically, as depicted in Fig. \ref{fig:tta-f3}(a), the AV-HuBERT upstream incorporates distinct front-end components—including an audio feature extractor and a video feature extractor—followed by a post-extraction projection, 12 Transformer layers, and interleaved LayerNorm layers. The intermediate outputs from these representations are integrated via learnable weights and subsequently decoded through a downstream two-layer BiLSTM for sequence recognition. This detailed hierarchy provides a granular structural basis for our subsequent layer-specific TTA configurations.

\subsubsection{ViSpeR.} In contrast, ViSpeR adopts a decoupled "modality-specific encoding + cross-modal fusion" approach, establishing a clear structural distinction from AV-SUPERB's joint upstream representation. As illustrated in Fig.~\ref{fig:tta-f3}(b), the audio stream is front-ended by a Conv1dResNet \cite{resnet1d_ori}, while a Conv3dResNet \cite{resnet3d_ori} extracts visual features from lip movements. Crucially, each modality is then independently processed through a dedicated deep encoder comprising 12 Conformer layers and 61 interleaved LayerNorm layers. Following cross-modal fusion, the integrated representations are decoded via two parallel branches: an autoregressive Transformer decoder (containing 6 Transformer layers and 19 LayerNorm layers) and a non-autoregressive Connectionist Temporal Classification (CTC) \cite{ctc} decoder. This highly modular, multi-layer configuration allows for specialized, modality-isolated feature refinement, providing explicit target layers for our subsequent block-wise and layer-specific TTA exploration.

\subsubsection{TTA Mechanism.} We employ SGEM \cite{SGEM_ori} as our core TTA mechanism. SGEM adapts model parameters in an unsupervised manner by minimizing a composite loss function comprising Generalized Entropy Minimization (GEM) and Negative Sampling (NS) components. The objective function is defined as:

\begin{equation}
\mathcal{L}_{\text{GEM}}=\frac{1}{L}\sum_{i=1}^{L}\frac{1}{1-\beta}\log\left(\sum_{j=1}^{C}p_{ij}^{\beta}\right),
\end{equation}
\begin{equation}
\mathcal{L}_{\text{NS}}=-\frac{1}{L}\sum_{i=1}^{L}\sum_{j=1}^{C}\mathbb{I}_{[p_{ij}^{\prime}<\tau]}\log(1-p_{ij}),
\end{equation}
\begin{equation}
\mathcal{L}_{\text{SGEM}}=\mathcal{L}_{\text{GEM}}+\lambda_{\text{NS}}\mathcal{L}_{\text{NS}},
\end{equation}
where $L$ denotes the number of speech frames, $C$ is the vocabulary size, and $p_{ij}$ is the probability of the $j$-th token at the $i$-th frame. The hyperparameters $\beta$, $\lambda_{\text{NS}}$, and $\tau$ dynamically regulate the adaptation sensitivity to the sequential outputs. Specifically, the entropic order $\beta > 1$ flattens the penalty curves of Rényi entropy, scaling the influence of high-confidence predictions to mitigate gradient vanishing under severe pathological domain shifts. The threshold $\tau$ acts as an unsupervised decision boundary, isolating low-confidence tokens whose probabilities fall below $\tau$ to explicitly penalize non-diagonal class confusion at every timestep. Lastly, the weight $\lambda_{\text{NS}}$ balances distribution sharpening and multi-class correlation reduction, effectively preventing the framework from degenerating into class collapse during episodic single-utterance adaptation.

The adaptation of the AVSR framework using SGEM follows a systematic pipeline:
\begin{enumerate}
\item \textbf{Initialization:} Load the pre-trained model and cache the initial parameters to ensure a clean state.
\item \textbf{Component selection:} Identify specific model components for adaptation and set them to a trainable state. In our study, we systematically evaluate individual or combined fine-tuning scopes to define the optimal adaptation framework, including LayerNorm parameters, audio/video feature extractors, specific sub-modules within the encoder or decoder infrastructure, and full-parameter adaptation of the entire architecture.
\item \textbf{Forward pass:} Process the sample to obtain output logits.
\item \textbf{Loss computation \& Update:} Calculate the  loss and update the trainable parameters via gradient descent.
\item \textbf{Iterative refinement:} Repeat Steps 3–4 for a fixed number of cycles to refine the adaptation.
\item \textbf{Parameter reset:} Upon completing the inference for a given sample, reset the model to its original pre-trained state using the cached parameters to avoid "catastrophic forgetting" or bias accumulation across samples.
\end{enumerate}

\section{Experiments}
\subsection{Datasets}

\noindent\textbf{CN-CVS}\footnote{\url{http://cncvs.cslt.org}} is a large-scale Mandarin Chinese audio-visual corpus comprising clips from broadcast news and internet speech shows. It consists of two subsets:
\begin{itemize}
    \item News: includes approximately 13,000 segments from 28 speakers (35 hours), characterized by constrained speaking styles and frontal camera angles;
    \item Speech: contains over 193,000 utterances from 2,500+ speakers (273 hours), featuring more natural prosody and varied head poses.
\end{itemize}

\noindent\textbf{CDSD}\footnote{\url{http://melab.psych.ac.cn/CDSD.html}} is a Chinese dysarthric speech database. For our experimental framework and cross-validation protocols, we curate an 8-hour target subset from 8 distinct speakers, with each speaker contributing approximately one hour of recordings. The detailed duration and quantity statistics across these subjects are systematically summarized in Table \ref{tab:cdsd data}. We specifically select subjects who simultaneously possess both raw, unsegmented video recordings and corresponding annotated sentence-level speech segments. The chosen videos are subsequently processed using our proposed source speech search-segmentation pipeline (Section \ref{sec:Source speech search-segmentation}), where the localized lip regions are cropped to facilitate visual feature extraction.

\subsection{Experimental Setup}
\label{sec:expset}

\subsubsection{DAVSR Training Configurations.}
We implemented our DAVSR models within the AV-SUPERB framework\footnote{\url{https://github.com/roger-tseng/av-superb/tree/main/downstream_tasks/av_asr}}, utilizing a pre-trained AV-HuBERT backbone\footnote{\url{https://dl.fbaipublicfiles.com/avhubert/model/lrs3/clean-pretrain/base_lrs3_iter5.pt}} as the upstream feature extractor and a two-layer BiLSTM for downstream decoding. We utilized the AV-HuBERT model pre-trained on LRS3 \cite{lrs3}. Despite the cross-lingual gap, this large-scale pre-training provides a robust foundation for extracting phonetic and visemic features. Recent work \cite{avhubertpth} has validated that AV-HuBERT effectively transfers to Mandarin tasks, demonstrating its capability to capture articulatory patterns across different languages. The character dictionary for our Chinese task comprises 5,536 characters. We evaluated three distinct training setups:
\begin{itemize}
    \item \textbf{AS-1 (LOSO):} A Leave-One-Speaker-Out (LOSO) scheme trained exclusively on the sentence-level aligned CDSD subset (denoted as "Aligned Data" in Table \ref{tab:cdsd data}). In each fold, the model is trained on seven speakers and evaluated on the remaining held-out speaker.
    \item \textbf{AS-2 (Cross-domain):} Zero-shot evaluation with the model trained entirely on the out-of-domain CN-CVS dataset.
    \item \textbf{AS-3 (Fine-tuned):} The AS-2 model is further fine-tuned on the aligned CDSD data following the same LOSO protocol.
\end{itemize}
To ensure convergence, training was terminated if the training loss stagnated for 10 consecutive epochs. The checkpoint yielding the minimum training loss was selected for subsequent testing and TTA experiments.

\subsubsection{TTA Setup.} We conduct our TTA experiments using both the AV-SUPERB and ViSpeR frameworks. For AV-SUPERB, the adaptation is initialized using the model checkpoints obtained from the three training configurations (AS-1, AS-2, and AS-3). For ViSpeR, TTA is performed on its pre-trained multilingual model\footnote{\url{https://github.com/m-burchi/ViSpeR}}, which was trained on a massive 5,412-hour multilingual corpus covering English, Mandarin, Arabic, Spanish, and French, with 787 hours specifically dedicated to Mandarin. This extensive pre-training provides a robust multilingual feature space and utilizes a vocabulary of 21,000 tokens to accommodate its broad linguistic coverage.

For all TTA configurations, we employ the AdamW optimizer with a cosine annealing scheduler, denoting the initial and terminal learning rates as $\eta_s$ and $\eta_e$, respectively. We fix the temperature and threshold at $(T,\tau) = (2.5, 0.4/C)$. To ensure strict experimental consistency, we adopt the baseline hyperparameter configurations validated by previous SGEM works. Furthermore, our ablation experiments systematically focus on task-specific component selection scopes (as summarized in Table \ref{tab:cdsd CER tta-ablation}) rather than exhaustive hyperparameter tuning, since exploring modular component combinations provides clearer structural insights into multimodal adaptation dynamics. To investigate the impact of specific architectural components, we define four adaptation scopes:
\begin{itemize}
    \item \textbf{LayerNorm:} Updating exclusively the LayerNorm parameters across the sub-structures ($\eta_{s}=2\times10^{-4}, \eta_{e}=2\times10^{-5}, \beta=1.5, \lambda_{\text{NS}}=1$).
    \item \textbf{Feature:} Restricting gradient propagation to the front-end features extractors, specifically targeting the Audio Feature Extractor and Video Feature Extractor blocks ($\eta_{s}=4\times10^{-5}, \eta_{e}=2\times10^{-5}, \beta=1.5, \lambda_{\text{NS}}=1$).
    \item \textbf{Module-Specific:} Localizing updates to sequence-level transduction blocks. For AV-SUPERB, this evaluates independent encoder or decoder blocks. For ViSpeR's decoupled streams, this fine-tunes the standalone audio encoder, video encoder, transformer decoder, or the CTC branch ($\eta_{s}=4\times10^{-6}, \eta_{e}=2\times10^{-6}, \beta=1.25, \lambda_{\text{NS}}=0.5$).
    \item \textbf{All:} Setting all architectural parameters across the entire multi-modal network to a trainable state simultaneously ($\eta_{s}=4\times10^{-7}, \eta_{e}=2\times10^{-7}, \beta=1.25, \lambda_{\text{NS}}=0.5$).
\end{itemize}

\begin{figure}[t]
\setlength{\abovecaptionskip}{0.1cm} 
\centerline{\includegraphics[width=11cm]{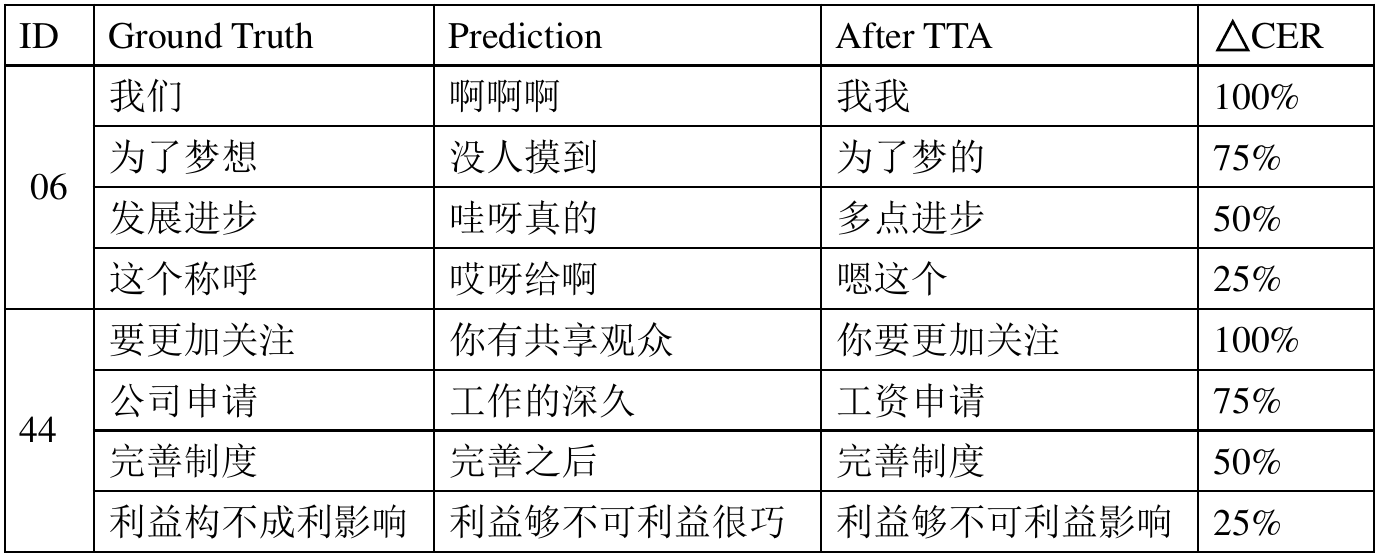}}
\caption{Qualitative examples of speech recognition transcripts from speakers 06 and 44 under their optimal model (ViSpeR), demonstrating the error-correction capability of the TTA.$\Delta\text{CER}$ quantifies the absolute reduction in CER.}
\label{fig:cn-case}
\end{figure}

\begin{figure*}[th]
\setlength{\abovecaptionskip}{0.1cm} 
\centerline{\includegraphics[width=13cm]{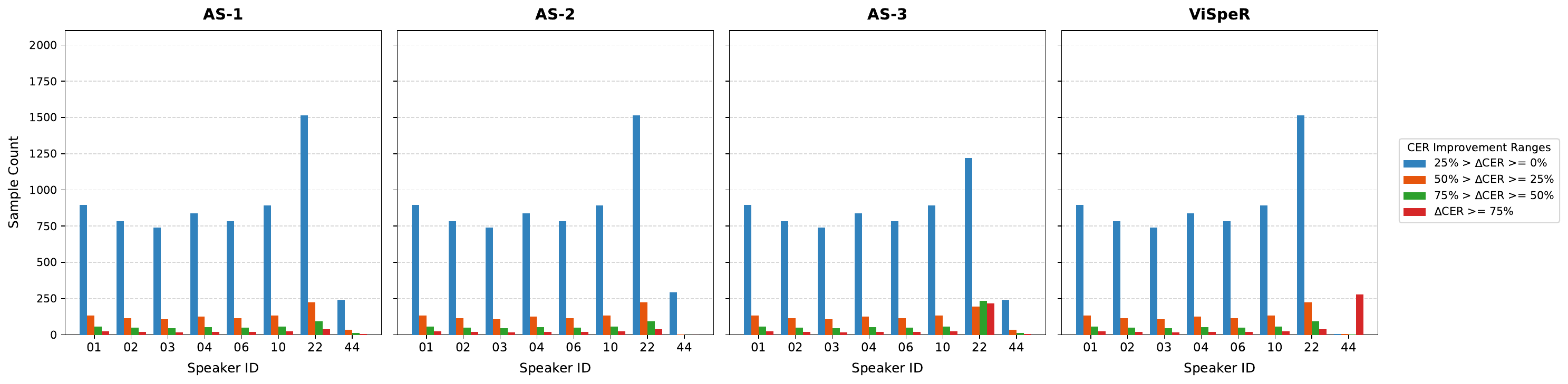}}
\caption{Distribution of testing utterances per speaker, categorized by the level of CER reduction achieved through TTA across four distinct evaluation frameworks: AS-1, AS-2, AS-3, and ViSpeR. To compensate for visual limitations, the samples are binned into four precise CER improvement intervals ($\Delta\text{CER} = \text{CER}_{\text{unadapted}} - \text{CER}_{\text{TTA}}$): marginal-to-low correction ($25\% > \Delta\text{CER} \ge 0\%$, blue bars), low-to-moderate correction ($50\% > \Delta\text{CER} \ge 25\%$, orange bars), moderate-to-high correction ($75\% > \Delta\text{CER} \ge 50\%$, green bars), and significant correction ($\Delta\text{CER} \ge 75\%$, red bars).}
\label{fig:f3}
\end{figure*}
\subsection{Results and Discussions}
\begin{table}[!t] 
    \setlength{\abovecaptionskip}{0.1cm} 
    \caption{TTA Performance (CER, \%) on unseen speakers across various adaptation iterations. The absolute CER reduction ($\Delta$) is computed as the baseline error minus the optimal adapted error ($\text{CER}_{\text{Baseline}} - \text{CER}_{\text{Optimal}}$), with the latter denoted by the bolded values in each row. The asterisk ($*$) highlights the best recognition performance achieved per speaker across all evaluated configurations.}
    \vspace{-6pt}
    \label{tab:cdsd CER tta-all} 
    \centering
    \begin{tabular}{c | l | c | c c c c | c}
        \hline 
        \textbf{Speaker ID} & \textbf{Model} & \textbf{Baseline} & \textbf{TTA-1} & \textbf{TTA-3} & \textbf{TTA-5} & \textbf{TTA-10} & \boldmath$\Delta$ \\
        \hline
        & \textbf{AS-1} & 88.4 & 87.6 & \textbf{87.1} & 87.8 & 87.7 & 1.3 \\
        \textbf{01} & \textbf{AS-2} & 74.4 & 72.1 & \textbf{71.0} & 71.7 & 72.2 & 3.4 \\
        & \textbf{AS-3} & 75.0 & 73.0 & \textbf{72.5} & 72.7 & 73.3 & 2.5 \\
        & \textbf{ViSpeR} & 58.9 & 55.7 & 48.1 & 47.3 & \textbf{47.1*} & \textbf{11.8} \\
        \hline
        & \textbf{AS-1} & 85.2 & 85.5 & \textbf{84.3} & 84.3 & 84.4 & 0.9 \\
        \textbf{02} & \textbf{AS-2} & 88.2 & 87.6 & \textbf{87.4} & 87.4 & 88.1 & 0.8 \\ 
        & \textbf{AS-3} & 75.8 & 75.9 & \textbf{75.6*} & 76.2 & 76.3 & 0.2 \\
        & \textbf{ViSpeR} & 83.1 & 80.3 & 78.0 & 78.1 & \textbf{76.9} & \textbf{6.2} \\
        \hline
        & \textbf{AS-1} & 88.1 & 89.1 & \textbf{88.7} &  89.2 & 88.9 & -0.6 \\
        \textbf{03} & \textbf{AS-2} & 86.4 &  87.0 & \textbf{86.5} & 88.0 & 87.7 & -0.1 \\
        & \textbf{AS-3} & 78.0  & 79.9 & \textbf{77.3} & 78.2 & 78.5 & 0.7 \\
        & \textbf{ViSpeR} & 74.5 & 72.6 & 68.3 & 68.4 & \textbf{66.7*} & \textbf{7.8} \\
        \hline
        & \textbf{AS-1} & 79.4 & 79.5 & \textbf{79.0} & 80.5 & 80.3 & 0.4 \\
        \textbf{04} & \textbf{AS-2} & 82.9 &  82.5 & \textbf{81.7} & 81.8 & 82.0 & 1.2 \\ 
        & \textbf{AS-3} & 57.9 & 57.8 & \textbf{56.8} & 57.5 & 57.1 & 1.1\\
        & \textbf{ViSpeR} & 57.4 & 57.6 & 55.3 & 53.4 & \textbf{53.0*} & \textbf{4.4}\\
        \hline
        & \textbf{AS-1} & 103.2 & 102.1 & \textbf{100.3} & 100.5 & 100.5 & 2.9 \\
        \textbf{06} & \textbf{AS-2} & 100.1 & 99.4 & \textbf{99.2} & 99.3 & 1.0 & 0.9 \\
        & \textbf{AS-3} & 98.6 & 98.1 & \textbf{97.7*} & 98.1 & 98.5 & 0.9 \\
        & \textbf{ViSpeR} & 139.3 & 129.0 &	102.8 & 99.5 & \textbf{97.9} & \textbf{41.4}\\
        \hline
        & \textbf{AS-1} & 92.1 & 92.3 & \textbf{91.5} & 92.2 & 92.8 & 0.6 \\
        \textbf{10} & \textbf{AS-2} & 92.3 & 92.3 & \textbf{92.2} & 92.5 & 92.7 & 0.1 \\
        & \textbf{AS-3} & 87.6 & 86.5 & \textbf{86.5*} & 86.7 & 86.7 & 1.1 \\
        & \textbf{ViSpeR} & 96.7 & 92.4 & 91.0 & 90.5 & \textbf{90.3} & \textbf{6.4}\\
        \hline
        & \textbf{AS-1} & 68.5 & 69.7 & \textbf{68.5} & 69.3 & 69.8 & 0.0 \\
        \textbf{22} & \textbf{AS-2} & 71.2 & 70.6 & \textbf{68.6} & 69.9 & 69.2 & \textbf{2.6} \\
        & \textbf{AS-3} & 56.5 & 56.4 & \textbf{54.0} & 56.1 & 54.7 & 2.5 \\
        & \textbf{ViSpeR} & 48.8 & 50.4 & 49.1 & 48.3 & \textbf{47.9*} & 0.9\\
        \hline
        & \textbf{AS-1} & 90.1 & 92.6 & \textbf{91.0} & 94.1 & 91.7 & -0.9 \\
        \textbf{44} & \textbf{AS-2} & 101.4 & 101.1 & \textbf{100.6} & 101.0 & 102.2 & 0.8 \\
        & \textbf{AS-3} & 87.3 & 87.0 & \textbf{86.7*} & 87.8 & 86.7 & 0.6 \\
        & \textbf{ViSpeR} & 146.5 & 114.8 & 100.0 & 99.9 & \textbf{99.6} & \textbf{46.9}\\
    \hline
    \end{tabular}
    \vspace{-10pt}
\end{table}

Table \ref{tab:cdsd CER tta-all} summarizes the DAVSR performance on unseen speakers before and after TTA. To ensure an accurate and fair interpretation of these results, it is essential to clarify that they are not directly comparable to the previous single-modal baseline results reported in the original CDSD study~\cite{cdsd_arxiv_ori}. While the original baseline leveraged a substantially larger volume of purely acoustic training speech (approximately 1 hour per speaker), our multimodal setting is strictly constrained by the limited availability of paired audio-visual data, yielding only about one-tenth of an hour of synchronized recordings per speaker. 

Table \ref{tab:cdsd CER tta-all} demonstrates that TTA consistently reduces the CER across most configurations, establishing a highly viable pathway for unsupervised model enhancement. While Speakers 03 and 44 exhibited marginal regressions under AS-1 and AS-2, the CER increases remained below 1\%, and TTA still provided net positive benefits for the most robust baseline model (AS-3). Speakers 06 and 44 initially presented exceptionally high CERs due to severe speech impairments, such as indistinct articulation and abnormal pauses. These severe impairments initially triggered model hallucinations, leading to meaningless outputs that further inflated the baseline error rates. However, as illustrated in Figure \ref{fig:cn-case}, TTA achieved noticeable improvements in prediction accuracy even in these challenging instances where the models initially generated semantically unrelated outputs.

By analyzing the experimental results across successive adaptation iterations, we can identify several key patterns and conclusions that correspond directly to the differing model architectures. As synthesized in Table \ref{tab:cdsd CER tta-all}, across the AV-SUPERB framework settings, the CER reductions typically plateaued or exhibited marginal fluctuations after three adaptation iterations. In contrast, the ViSpeR framework demonstrated a distinct and continuous error reduction trajectory that extended smoothly up to ten iterations without showing premature optimization saturation. This persistent self-learning progression is primarily facilitated by ViSpeR's extensive 5,412-hour multilingual pre-training, which endows the model with superior generalization capabilities and robust self-adaptation performance. Such highly generalized representations provide stable gradient trajectories during unsupervised entropy minimization, enabling the model to extract more granular adjustments from unlabeled target streams over extended cycles. To rigorously verify the reliability of these performance improvements, paired $t$-tests were executed to compare the unadapted baselines against the optimal configurations (specifically, Baseline vs. TTA-3 for AS-3, and Baseline vs. TTA-10 for ViSpeR). The evaluations yielded $p$-values strictly below 0.05 across both scenarios, confirming that the TTA-driven improvements are statistically significant and mathematically robust rather than stochastic variations.

\begin{table}[!t] 
    \setlength{\abovecaptionskip}{0.0cm} 
    \caption{The performance of different TTA settings for speaker 01 under various frameworks (CER, \%).}
    \vspace{-6pt}
    \label{tab:cdsd CER tta-ablation} 
    \centering
    \begin{tabular}{c | c | c | c c c c |c}
        \hline
        \textbf{Model} & \textbf{Setting} & \textbf{Baseline} &\textbf{TTA-1} & \textbf{TTA-3} & \textbf{TTA-5} & \textbf{TTA-10} & \boldmath$\Delta$ \\
        \hline 
        & \textbf{LayerNorm} & 88.4 & 87.7 & \textbf{87.6} & 87.7 & 88.5 & 0.8\\
        & \textbf{Feature} & 88.4 & 87.6 & \textbf{87.1} & 87.8 & 87.7 & \textbf{1.3}\\
    \textbf{AS-1} & \textbf{Encoder} & 88.4 & 87.7 & \textbf{87.5} & 87.9 & 88.7 & 0.9\\
        & \textbf{Decoder} & 88.4 & 87.8 & \textbf{87.6} & 87.9 & 88.7 & 0.8\\
        & \textbf{All} & 88.4 & \textbf{87.6} & \textbf{87.6} & 87.7 & 88.2 & 0.8\\
        \hline
        & \textbf{LayerNorm} & 74.4 & 72.4 & \textbf{72.1} & 73.1 & 73.2 & 2.3 \\
        & \textbf{Feature} & 74.4 & 72.1 & \textbf{71.0} &	71.7 & 72.2 & \textbf{3.4} \\     
   \textbf{AS-2} & \textbf{Encoder} & 74.4 & 72.4 & \textbf{72.0} & 73.1 & 73.1 & 2.4\\
        & \textbf{Decoder} & 74.4 & 72.7 & \textbf{72.3} & 73.5 & 74.1 & 2.1 \\     
        & \textbf{All} & 74.4 & 72.7 & \textbf{72.3} & 73.4 & 73.8 & 2.1\\
        \hline
        & \textbf{LayerNorm} & 75.0 & 73.4 & \textbf{73.0} & 73.5 & 73.8 & 2.0 \\
        & \textbf{Feature} & 75.0 & 73.3 &	\textbf{72.5} & 73.4 & 73.3 & \textbf{2.5} \\
    \textbf{AS-3} & \textbf{Encoder} & 75.0 & 73.3 & \textbf{72.9} & 73.4 &	73.8 & 2.1 \\
        & \textbf{Decoder} & 75.0 & 73.3 & \textbf{73.2} & 73.7 & 74.3 & 1.8 \\
        & \textbf{All} & 75.0 & 73.4 & \textbf{73.2} & 73.6 & 74.2 & 1.8 \\
    \hline
        & \textbf{LayerNorm} & 58.9	& 50.3 & \textbf{49.4} & 50.7 & 54.6 & 9.5 \\
        & \textbf{Feature} & 58.9 & 58.9 & 58.1 & 56.7 & \textbf{56.1} & 2.8 \\
        & \textbf{Audio Encoder} & 58.9 & 52.4 & 49.4 &	\textbf{49.0} & 51.4 & 9.9 \\
    \textbf{ViSpeR} & \textbf{Video Encoder} & 58.9 & 54.2 & 52.8 &	\textbf{52.3} & 54.3 & 6.6 \\
        & \textbf{Transformer Decoder} & 58.9 & \textbf{59.6} & 60.4 & 60.5 & 60.5 & -0.7 \\
        & \textbf{CTC Decoder} & 58.9 & \textbf{58.9} & 58.9 & 58.9 & 58.9 & 0.0 \\
        & \textbf{All} & 58.9 & 55.7 & 48.1 & 47.3 &	\textbf{47.1} & \textbf{11.8} \\
    \hline
    \end{tabular}
    \vspace{-10pt}
\end{table}

Table \ref{tab:cdsd CER tta-ablation} details the ablation results of different model components under various TTA configurations. Using Speaker 01 as a representative scenario, the two frameworks exhibit distinct optimal adaptation configurations, reflecting their different underlying architectures. To maintain a concise presentation, the experimental results for the other seven speakers are omitted from the table, as they exhibit identical trends and lead to the same architectural conclusions as Speaker 01.

Within the AV-SUPERB framework, the optimal TTA parameter scope is highly consistent across all three training configurations (AS-1, AS-2, and AS-3). Specifically, AV-SUPERB consistently achieves its lowest CER by adapting the upstream front-end feature extractors (the "Feature" setup). Conversely, updating the downstream sequence decoder (whether individually as the "Decoder" setup or collectively within the "All" setup) yields the least optimal improvements. This behavior indicates that for representation-based modular frameworks like AV-SUPERB, propagating unsupervised gradients into the downstream recurrent decoder introduces optimization noise and over-parameterization. Restricting the self-learning adjustment to the representation space ("Feature") successfully adapts the front-end to individual domain shifts while preserving the stable linguistic decoding capacity of the downstream BiLSTM.

In contrast, the ViSpeR framework exhibits completely different structural requirements. While partial updates, such as adapting only the LayerNorm parameters or individual encoder streams, provide notable error reductions compared to the unadapted baseline, ViSpeR achieves its absolute optimal performance (yielding a net reduction of 11.8\%) when all architectural parameters are entirely unfrozen ("All" setup). Crucially, modifying only the Transformer Decoder or the CTC Decoder yields negligible improvements or, in the case of Transformer Decoder, a marginal regression of -0.7\%. This phenomenon suggests that for tightly coupled end-to-end architectures like ViSpeR, the modality-specific encoders and downstream joint-space cross-attention pathways are deeply interdependent. Unfreezing all parameters prevents representation mismatch between layers, enabling the multi-modal ResNet front-ends and downstream decoding networks to cooperatively recalibrate their alignment weights based on the unsupervised target criteria.

To investigate the granularity of these performance gains at the utterance level, Figure \ref{fig:f3} illustrates the distribution of sample-specific CER reductions across all eight speakers under the four evaluation frameworks. For most speakers (e.g., Speakers 01, 02, 03, 04, 06, and 10), the vast majority of utterances fall within the marginal-to-low improvement tier ($25\% > \Delta\text{CER} \ge 0\%$, blue bars). This indicates that for speakers whose unadapted baseline recognition was already relatively stable, TTA primarily delivers highly reliable, small-scale refinements.

In contrast, speakers presenting more complex pathological speech patterns exhibit higher-magnitude improvements. For Speaker 22, a substantial portion of samples are successfully shifted into the moderate-to-high and significant correction tiers. This is particularly evident under the AS-3 fine-tuned framework, where both the green ($75\% > \Delta\text{CER} \ge 50\%$) and red ($\Delta\text{CER} \ge 75\%$) bars exceed 200 samples (with the green bar reaching nearly 240).

The most dramatic architectural and sample-level contrast is observed in Speaker 44, who initially suffered from exceptionally severe speech impairments. Under the modular AV-SUPERB configurations (AS-1, AS-2, and AS-3), Speaker 44 exhibits almost exclusively low-tier or negligible improvements, with the utterances almost entirely concentrated in the blue bar ($25\% > \Delta\text{CER} \ge 0\%$). Under the zero-shot AS-2 framework, this manifests as a single, exclusive blue bar representing the entire 294-sample set. However, under the ViSpeR framework, Speaker 44 demonstrates a sample-level transition: the low-tier blue bar is virtually eliminated, replaced instead by a dominant red bar ($\Delta\text{CER} \ge 75\%$) containing nearly 280 samples. This highly localized, massive correction underscores the powerful synergy between ViSpeR's highly generalized multilingual foundation and our TTA strategy.

\section{Conclusions}
This work investigates the application of TTA to sentence-level DAVSR. To facilitate this study under severe data scarcity, we construct a structured multimodal corpus from the raw CDSD database using our proposed "source speech search-segmentation" pipeline. Our evaluations across both AV-SUPERB and ViSpeR frameworks suggest that TTA can partially alleviate performance degradation for unseen pathological speakers by leveraging unlabeled test data. Although the improvements are constrained by the inherent heterogeneity of dysarthric speech, these findings highlight TTA as a promising auxiliary approach for enhancing model generalizability.

\bibliographystyle{splncs04}
\bibliography{mybib}

\begin{thebibliography}{10}
\providecommand{\url}[1]{\texttt{#1}}
\providecommand{\urlprefix}{URL }
\providecommand{\doi}[1]{https://doi.org/#1}

\bibitem{lrs3}
Afouras, T., Chung, J.S., Zisserman, A.: Lrs3-ted: a large-scale dataset for visual speech recognition. arXiv preprint arXiv:1809.00496  (2018)

\bibitem{davsr20241_ori}
Chen, X., Wang, Y., Wu, X., Wang, D., Wu, Z., Liu, X., Meng, H.: Exploiting audio-visual features with pretrained av-hubert for multi-modal dysarthric speech reconstruction. In: ICASSP 2024-2024 IEEE International Conference on Acoustics, Speech and Signal Processing (ICASSP). pp. 12341--12345. IEEE (2024)

\bibitem{davsr20242_ori}
Chen, X., Yang, D., Wang, D., Wu, X., Wu, Z., Meng, H.: Colm-dsr: Leveraging neural codec language modeling for multi-modal dysarthric speech reconstruction. arXiv preprint arXiv:2406.08336  (2024)

\bibitem{avhubertpth}
Chien, Y.L., Chen, H.H., Yen, M.C., Tsai, S.W., Wang, H.M., Tsao, Y., Chi, T.S.: Audio-visual mandarin electrolaryngeal speech voice conversion. arXiv preprint arXiv:2306.06652  (2023)

\bibitem{ctc}
Graves, A., Fern{\'a}ndez, S., Gomez, F., Schmidhuber, J.: Connectionist temporal classification: labelling unsegmented sequence data with recurrent neural networks. In: Proceedings of the 23rd international conference on Machine learning. pp. 369--376 (2006)

\bibitem{resnet1d_ori}
He, K., Zhang, X., Ren, S., Sun, J.: Deep residual learning for image recognition. In: Proceedings of the IEEE conference on computer vision and pattern recognition. pp. 770--778 (2016)

\bibitem{davsr20192_ori}
Hu, S., Liu, S., Chang, H.F., Geng, M., Chen, J., Chung, L.W., Hei, T.K., Yu, J., Wong, K.H., Liu, X., et~al.: The cuhk dysarthric speech recognition systems for english and cantonese. In: INTERSPEECH. pp. 3669--3670 (2019)

\bibitem{sita_ori}
Khurana, A., Paul, S., Rai, P., Biswas, S., Aggarwal, G.: Sita: Single image test-time adaptation. arXiv preprint arXiv:2112.02355  (2021)

\bibitem{SGEM_ori}
Kim, C., Park, J., Shim, H., Yang, E.: Sgem: Test-time adaptation for automatic speech recognition via sequential-level generalized entropy minimization. arXiv preprint arXiv:2306.01981  (2023)

\bibitem{uaspeech_ori}
Kim, H., Hasegawa-Johnson, M., Perlman, A., Gunderson, J., Huang, T.S., Watkin, K., Frame, S.: Dysarthric speech database for universal access research. In: Ninth Annual Conference of the International Speech Communication Association (2008)

\bibitem{SUTA_ori}
Lin, G.T., Li, S.W., Lee, H.y.: Listen, adapt, better wer: Source-free single-utterance test-time adaptation for automatic speech recognition. arXiv preprint arXiv:2203.14222  (2022)

\bibitem{davsr20191_ori}
Liu, S., Hu, S., Wang, Y., Yu, J., Su, R., Liu, X., Meng, H.: Exploiting visual features using bayesian gated neural networks for disordered speech recognition. In: INTERSPEECH. pp. 4120--4124 (2019)

\bibitem{davsr2020_ori}
Liu, S., Xie, X., Yu, J., Hu, S., Geng, M., Su, R., Zhang, S.X., Liu, X., Meng, H.: Exploiting cross-domain visual feature generation for disordered speech recognition. In: Interspeech. pp. 711--715 (2020)

\bibitem{autoavsr_ori}
Ma, P., Haliassos, A., Fernandez-Lopez, A., Chen, H., Petridis, S., Pantic, M.: Auto-avsr: Audio-visual speech recognition with automatic labels. In: ICASSP 2023-2023 IEEE International Conference on Acoustics, Speech and Signal Processing (ICASSP). pp.~1--5. IEEE (2023)

\bibitem{ViSpeR_ori}
Narayan, S., Djilali, Y.A.D., Singh, A., Bihan, E.L., Hacid, H.: Visper: Multilingual audio-visual speech recognition. arXiv preprint arXiv:2406.00038  (2024)

\bibitem{avhubert_ori}
Shi, B., Hsu, W.N., Lakhotia, K., Mohamed, A.: Learning audio-visual speech representation by masked multimodal cluster prediction. arXiv preprint arXiv:2201.02184  (2022)

\bibitem{resnet3d_ori}
Stafylakis, T., Tzimiropoulos, G.: Combining residual networks with lstms for lipreading. arXiv preprint arXiv:1703.04105  (2017)

\bibitem{cdsd_arxiv_ori}
Sun, M., Gao, M., Kang, X., Wang, S., Du, J., Yao, D., Wang, S.J.: Cdsd: Chinese dysarthria speech database. arXiv preprint arXiv:2310.15930  (2023)

\bibitem{lhuc_ori}
Swietojanski, P., Li, J., Renals, S.: Learning hidden unit contributions for unsupervised acoustic model adaptation. IEEE/ACM Transactions on Audio, Speech, and Language Processing  \textbf{24}(8),  1450--1463 (2016)

\bibitem{var_ori}
Tomanek, K., Seaver, K., Jiang, P.P., Cave, R., Harrell, L., Green, J.R.: An analysis of degenerating speech due to progressive dysarthria on asr performance. In: ICASSP 2023-2023 IEEE International Conference on Acoustics, Speech and Signal Processing (ICASSP). pp.~1--5. IEEE (2023)

\bibitem{avsuperb_ori}
Tseng, Y., Berry, L., Chen, Y.T., Chiu, I.H., Lin, H.H., Liu, M., Peng, P., Shih, Y.J., Wang, H.Y., Wu, H., et~al.: Av-superb: A multi-task evaluation benchmark for audio-visual representation models. In: ICASSP 2024-2024 IEEE International Conference on Acoustics, Speech and Signal Processing (ICASSP). pp. 6890--6894. IEEE (2024)

\bibitem{tent_ori}
Wang, D., Shelhamer, E., Liu, S., Olshausen, B., Darrell, T.: Tent: Fully test-time adaptation by entropy minimization. arXiv preprint arXiv:2006.10726  (2020)

\bibitem{pbdsr_ori}
Wang, S., Zhao, S., Zhou, J., Kong, A., Qin, Y.: Enhancing dysarthric speech recognition for unseen speakers via prototype-based adaptation. arXiv preprint arXiv:2407.18461  (2024)

\bibitem{dda}
Wang, S., Zhao, S., Zhou, J., Qin, Y.: Zero-and one-shot data augmentation for sentence-level dysarthric speech recognition in constrained scenarios. arXiv preprint arXiv:2510.16700  (2025)

\bibitem{pblrdwws_ori}
Wang, S., Zhou, J., Zhao, S., Qin, Y.: Pb-lrdwws system for the slt 2024 low-resource dysarthria wake-up word spotting challenge. In: 2024 IEEE Spoken Language Technology Workshop (SLT). pp. 586--591. IEEE (2024)

\bibitem{sapc_ori}
Wang, S., Zhou, J., Zhao, S., Qin, Y.: A self-training approach for whisper to enhance long dysarthric speech recognition. arXiv preprint arXiv:2506.22810  (2025)

\bibitem{davsr2023_ori}
Yu, C., Su, X., Qian, Z.: Multi-stage audio-visual fusion for dysarthric speech recognition with pre-trained models. IEEE Transactions on Neural Systems and Rehabilitation Engineering  \textbf{31},  1912--1921 (2023)

\end{thebibliography}

\end{document}